# Quantum Čerenkov Effect from Hot Carriers in Graphene: An Efficient Plasmonic Source

Ido Kaminer[1], Yaniv Tenenbaum Katan[2], Hrvoje Buljan[3], Yichen Shen[1], Ognjen Ilic[1], Josué J. López[1], Liang Jie Wong[4], John D. Joannopoulos[1], and Marin Soljačić[1]

[1]Department of Physics, Massachusetts Institute of Technology,
77 Massachusetts Avenue, Cambridge 02139, Massachusetts, USA

[2]Physics Department and Solid State Institute, Technion, Haifa 32000, Israel

[3]Department of Physics, University of Zagreb, Bijenička c. 32, 10000 Zagreb, Croatia

[4]Singapore Institute of Manufacturing Technology, 71 Nanyang Drive, Singapore 638075

**Graphene plasmons (GPs) have been found to be an exciting plasmonic platform, thanks to their high field confinement and low phase velocity, motivating contemporary research to revisit established concepts in light-matter interaction. In a conceptual breakthrough that is now more than 80 years old, Čerenkov showed how charged particles emit shockwaves of light when moving faster than the phase velocity of light in a medium. To modern eyes, the Čerenkov effect (ČE) offers a direct and ultrafast energy conversion scheme from charge particles to photons. The requirement for relativistic particles, however, makes ČE-emission inaccessible to most nanoscale electronic and photonic devices. We show that GPs provide the means to overcome this limitation through their low phase velocity and high field confinement. The interaction between the charge carriers flowing inside graphene and GPs presents a highly efficient 2D Čerenkov emission, giving a versatile, tunable, and ultrafast conversion mechanism from electrical signal to plasmonic excitation.**

Achieving ultrafast conversion of electrical to optical signals at the nanoscale using plasmonics [1,2] is a long-standing goal, due to its potential to revolutionize electronics and allow ultrafast communication and signal processing. Plasmonic systems combine the benefits of high frequencies ($10^{14}$-$10^{15}$ Hz) with those of small spatial scales, thus avoiding the limitation of conventional photonic systems, by using the strong field confinement of plasmons. However, the realization of plasmonic sources that are electrically pumped, power efficient, and compatible with current device fabrication processes (e.g. CMOS), is a formidable challenge. In recent years, several groups have demonstrated the potential of surface plasmons as a platform for strong and ultrafast light-matter interaction [3-6]. Graphene's tunability and strong field confinement [7-10] have motivated proposals for the use of GPs [7,11-13] in electrically-pumped plasmonic sources [14] and in the conversion of electrical energy into luminescence [15-17].

Here we show that under proper conditions charge carriers propagating *in* graphene can efficiently excite GPs, through a process that can be understood as 2D Čerenkov emission. Graphene provides a platform in which the flow of charge alone is sufficient for Čerenkov radiation, eliminating the need for accelerated charge particles in vacuum chambers, and opening up a new platform for the study of ČE and its applications, especially as a novel plasmonic source. On one hand, hot charge carriers moving with high velocities (up to the Fermi velocity $v_f \approx 10^6 \frac{m}{s}$) are considered possible, even in relatively large sheets of graphene ($10 \mu m$ and more [18]). On the other hand, plasmons in graphene can have an exceptionally slow phase velocity, down to a few hundred times slower than the speed of light [7,9,19]. This creates a scenario where velocity matching between charge carriers and plasmons is possible, enabling the emission of GPs from electrical excitations (hot carriers) at very high rates. This paves the way to new devices utilizing the ČE on the nanoscale, a prospect made even more attractive by the dynamic tunability of the Fermi level of graphene. For a wide range of parameters, the emission rate of GPs is significantly higher than that of photons or phonons, suggesting that taking advantage of the ČE increases the efficiency of energy conversion from electrical energy to plasmons, approaching 100%. We show that, contrary to expectations, plasmons can be created at energies above $2E_f$ – thus exceeding energies attainable by photon emission – resulting in a plasmon spectrum that extends from terahertz to near infrared frequencies and possibly into the visible range. Furthermore, we show that tuning the Fermi energy by external voltage can control the parameters (direction and frequency) of enhanced emission. This tunability also reveals regimes of backward GP emission, and regimes of forward GP emission with low angular spread, emphasizing the uniqueness of ČE from hot carriers flowing in graphene. Surprisingly, we find that GP emission can also result from intraband transitions that are made possible by plasmonic losses. These kinds of transitions can become significant, and might help explain several phenomena observed in graphene devices, such as current saturation [20], high frequency radiation spectrum from graphene [17,21], and the black body radiation spectrum that seems to relate to extraordinary high electron temperatures [22].

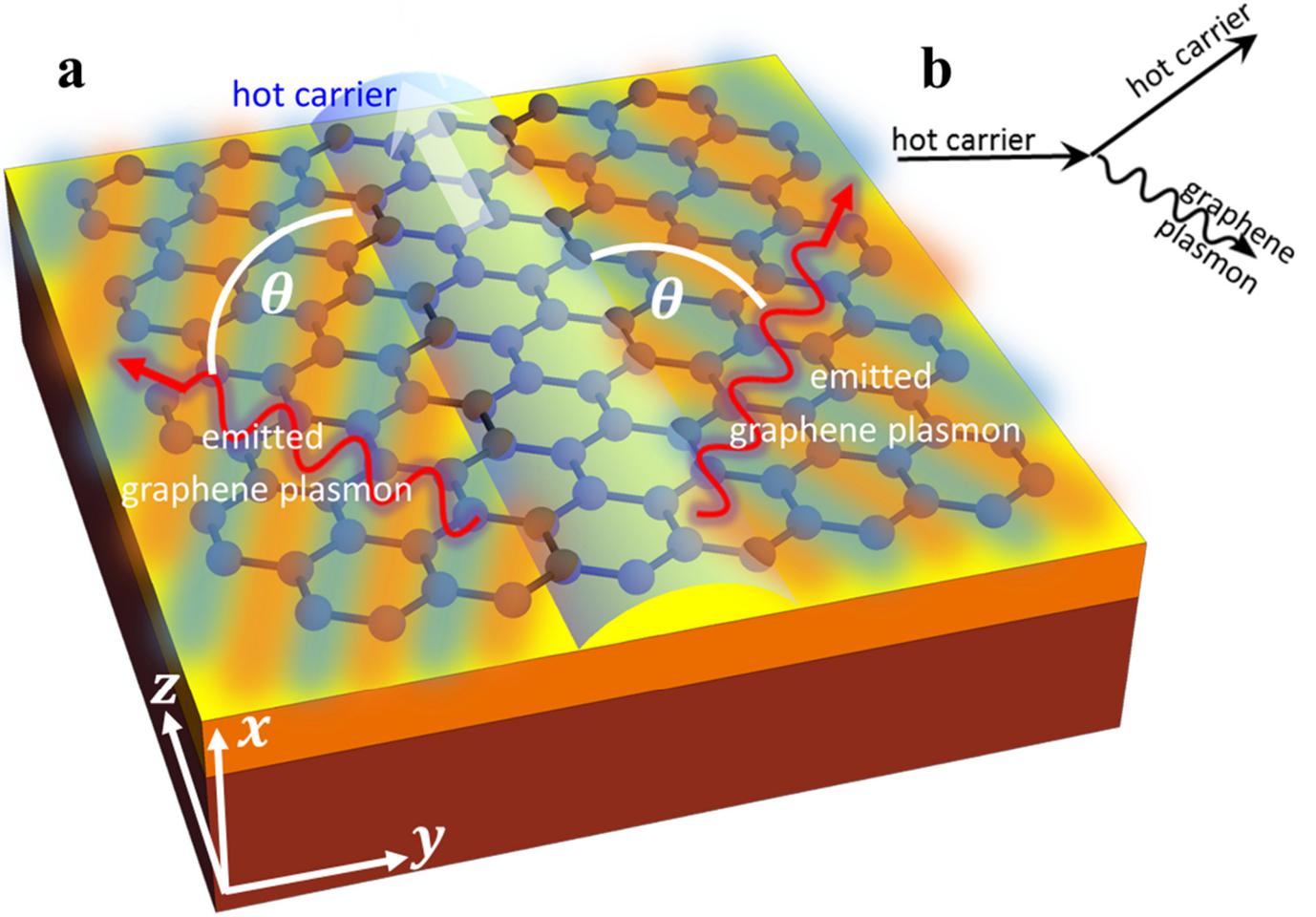

**Figure 1: Illustration of the plasmon emission from charge carriers in graphene via a 2D Čerenkov process.** (a) GP emission in graphene from a hot carrier flowing inside it. The Čerenkov angle into which the GPs are emitted is denoted by $\theta$. (b) A diagram describing the GP emission process from a hot carrier in graphene.

Recent studies [23-25], which focus on cases of classical free charge particles moving *outside* graphene, have revealed strong Čerenkov-related GP emission resulting from the charge particle-plasmon coupling. In contrast, in this work we focus on the study of charge carriers *inside* graphene, as illustrated in Fig.1a. For this purpose, we develop a quantum theory of ČE in graphene. As we shall see, our analysis of this system gives rise to a variety of novel Čerenkov-induced plasmonic phenomena. The conventional threshold of the ČE in either 2D or 3D ($v > v_p$) may seem unattainable for charge carriers *in* graphene, because they are limited by the Fermi velocity $v \leq v_f$, which is smaller than the GP phase velocity $v_f < v_p$, as shown by the random phase approximation calculations [19,26]. However, we show that quantum effects come into play to enable these charge carriers to surpass the actual ČE threshold. Specifically, the actual ČE threshold for free electrons is shifted from its classically-predicted value by the quantum recoil of electrons upon photon emission [27,28]. Because of this shift, the actual ČE velocity threshold can in fact lie below the velocity of charge carriers in graphene, contrary to the conventional predictions. At the core of the modification of the quantum ČE is the linearity of the charge carrier energy-momentum relation (Dirac cone). Consequently, a careful choice of parameters (e.g. Fermi energy, hot carrier energy) allows the ČE threshold to be attained – resulting in significant enhancements and high efficiencies of energy conversion from electrical to plasmonic excitation.

The quantum ČE can be described as a spontaneous emission process of a charge carrier emitting into GPs, calculated by Fermi's golden rule [27,29]. In our case the matrix elements must be obtained from the light-matter interaction term in the graphene Hamiltonian, illustrated by a diagram like Fig.1b. To model the GPs, we use the random phase approximation [19,26,30], combined with a frequency-dependent phenomenological lifetime [19] to account for additional loss mechanisms such as optical phonons and scattering from impurities in the sample (assuming graphene mobility of $\mu = 2000 \, cm^2/Vsec$). This approach has been shown to give good agreement with experimental results [8,12,13,31,32]. The graphene sheet is in the $yz$ plane, and the charge carrier is moving in the $z$ direction (Fig.1a). For the case of low-loss GPs, the calculation reduces to the following integral (Lossy GPs are described later in this work – Eq.4).

$$\Gamma = \frac{2\pi}{\hbar} \int_{-\infty}^{\infty} \left| M_{\mathbf{k}_i \to \mathbf{k}_f + \mathbf{q}} \right|^2 \delta\left( E_{\mathbf{k}_i} - \hbar\omega(\mathbf{q}) - E_{\mathbf{k}_f} \right) \frac{d^2\mathbf{q}}{(2\pi)^2/A} \frac{d^2\mathbf{k}_f}{(2\pi)^2/A} \quad (1a)$$

$$M_{\mathbf{k}_i \to \mathbf{k}_f + \mathbf{q}} = q_e (2\pi)^2 \delta(q_y + k_{fy}) \delta(k_{iz} - q_z - k_{fz}) v_f \sqrt{\frac{\hbar q}{\varepsilon_0 \widetilde{\omega}(\mathbf{q}) A^3}} \cdot [\text{SP}] \quad (1b)$$

Where $M_{\mathbf{k}_i \to \mathbf{k}_f + \mathbf{q}}$ is the matrix element, $A$ is the surface area used for normalization, $q_e$ is the electric charge, $\varepsilon_0$ is the vacuum permittivity, [SP] is the spinor-polarization coupling term, and $\widetilde{\omega}(\mathbf{q})$ is the GP dispersion-based energy normalization term [33] ($\widetilde{\omega}(\mathbf{q}) = \bar{\epsilon}_r \, \omega \cdot v_p/v_g$, using the group velocity $v_g = \partial \omega/\partial q$). The GP momentum $\mathbf{q} = (q_y, q_z)$ satisfies $\omega^2/v_p^2 = q_y^2 + q_z^2$, with the phase velocity $v_p = v_p(\omega)$ or $v_p(\mathbf{q})$ obtained from the plasmon dispersion relation as $v_p = \omega/q$. The momenta of the incoming (outgoing) charge carrier $\mathbf{k_i} = (k_{iy}, k_{iz})$ ($\mathbf{k_f} = (k_{fy}, k_{fz})$) correspond to energies $E_{\mathbf{k}_i}$ ($E_{\mathbf{k}_f}$) according to the conical momentum-energy relation $E_{\mathbf{k}}^2 = \hbar^2 v_f^2 (k_y^2 + k_z^2)$. The charge velocity is $v = E_{\mathbf{k}}/|\hbar \mathbf{k}|$, which equals a constant ($v_f$). The only approximation in Eq.1 and henceforth, comes from the standard assumption of high GP confinement (free space wavelength / GP wavelength >> 1) [19]. Substituting Eq.1a into Eq.1b we obtain (denoting $E_i = E_{\mathbf{k}_i}$):

$$\Gamma = \int_{-\infty}^{\infty} \frac{\alpha c \hbar \, v_g(\mathbf{q})}{\bar{\epsilon}_r \, v_p^2(\mathbf{q})/v_f^2} \delta(q_y + k_{fy}) \delta(k_{iz} - q_z - k_{fz}) \delta\left(E_i - \hbar\omega(\mathbf{q}) - E_{\mathbf{k}_f}\right) |SP|^2 d^2\mathbf{q} \, d^2\mathbf{k}_f \quad (2)$$

Where $\alpha$ ($\approx \frac{1}{137}$) is the fine structure constant, $c$ is the speed of light, and $\bar{\epsilon}_r$ is the relative substrate permittivity obtained by averaging the permittivity on both sides of the graphene. We assume $\bar{\epsilon}_r = 2.5$ for all the figures. Because material dispersion is neglected, all spectral features are uniquely attributed to the GP dispersion and its interaction with charge carriers and not to any frequency dependence of the dielectrics. We further define the angle $\varphi$ for the outgoing charge and $\theta$ for the GP, both relative to the z axis, which is the direction of the incoming charge. This notation allows us to simplify the spinor-polarization coupling term [SP] for charge carriers inside graphene to $|SP|^2 = \cos^2(\theta - \varphi/2)$ or $|SP|^2 = \sin^2(\theta - \varphi/2)$ for intraband or interband transitions respectively. The delta functions in Eq.2 restrict the emission to two angles $\theta = \pm\theta_{\check{C}}$ (a clear signature of the ČE), and so we simplify the rate of emission to:

$$\cos(\theta_{\check{C}}) = \frac{v_p}{v_f}\left[1 - \frac{\hbar\omega}{2E_i}\left(1 - \frac{v_f^2}{v_p^2}\right)\right] \quad (3a)$$

$$\Gamma_\omega = \frac{2\alpha c}{v_f \bar{\epsilon}_r} \frac{\left|1 - \frac{\hbar\omega}{2E_i}\left(1 + \frac{v_f}{v_p}\cos(\theta_{\check{C}})\right)\right|}{|\sin(\theta_{\check{C}})|} = \frac{2\alpha c}{v_f \bar{\epsilon}_r}\left|\frac{\sin(\theta_{\check{C}})}{1 - v_p^2/v_f^2}\right| \quad (3b)$$

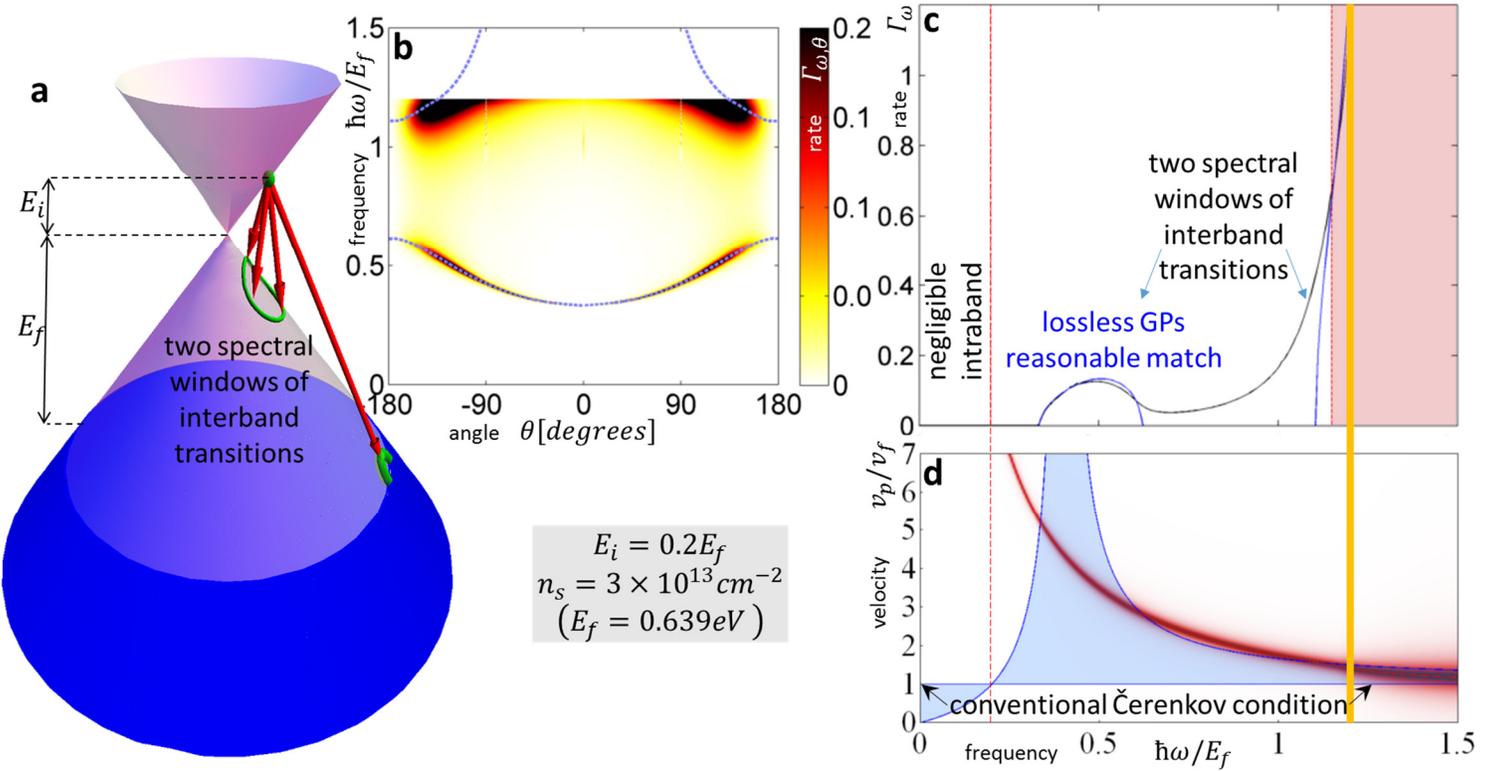

**Figure 2: GP emission from hot carriers.** (a) Illustration of the possible transitions. (b) Map of GP emission rate as a function of frequency and angle, Eq.4. We find most of the GP emission around the dashed blue curves that are exactly found by the Čerenkov angle Eq.3a. (c) Spectrum of the ČE GP emission process, with the red regime marking the area of high losses (as in [19]), the vertical dotted red line dividing between interband to intraband transitions, and the thick orange line marking the spectral cutoff due to the Fermi sea beyond which all states are occupied. Black – emission spectrum with GP losses, Eq.4. Blue – lossless emission approximation, Eq.3. (d) Explaining the GP emission with the quantum ČE. The red curve shows the GP phase velocity, with its thickness illustrating the GP loss. The blue regime shows the range of allowed velocities according to the quantum ČE. We find enhanced GP emission in the frequencies for which the red curve crosses the blue regime, either directly or due to the curve thickness. All figures are presented in normalized units except for the angle shown in degrees.

We note in passing that by setting $\hbar \to 0$ in the above expressions we recover the classical 2D ČE, including the Čerenkov angle $\cos(\theta_{\check{c}}) = v_p/v$, that can also be obtained from a purely classical electromagnetic calculation. However, while charge particles outside of graphene satisfy $\hbar\omega \ll E_i$, making the classical approximation almost always exact [27,28], the charges flowing inside graphene can have much lower energies because they are massless. Consequently, the introduced $\hbar$ terms in the ČE expression modifies the conventional velocity threshold significantly, allowing ČE to occur for lower charge velocities. e.g., while the conventional ČE requires charge velocity above the GP phase velocity ($v > v_p$), Eq.3a allows ČE below it, and specifically requires the velocity of charge carriers in graphene ($v = v_f$) to reside between $v_p > v_f > v_p \left|1 - \frac{2E_i}{\hbar\omega}\right|$. Physically, the latter case involves interband transitions made possible when graphene is properly doped: when the charge carriers are hot electrons (holes) interband ČE requires negatively (positively) doped graphene. Figures 2,3 demonstrate this interband ČE that indeed occurs for charge velocities below the conventional velocity threshold. More generally, the inequalities can be satisfied in two spectral windows simultaneously for the same charge carrier, due to the frequency dependence of the GP phase velocity (shown by the intersection of the red curve with the blue regime in Fig.2d). Moreover, part of the radiation (or even most of it, as in Fig.2) can be emitted backward, which is considered impossible for ČE in conventional materials [34,35]. Several spectral cutoffs appear in Figs.2c,3c,4c, as seen by the range of non-vanishing blue spectrum. These can be found by substituting $\theta_{\check{c}} = 0$ in Eq.3a, leading to $\hbar\omega_{cutoff} = 2E_i/(1 \pm v_f/v_p)$, exactly matching the points where the red curve in Figs.2d,3d,4d crosses the border of the blue regime. The upper most frequency cutoff marked by the thick orange line in Figs.2-4 occurs at $\hbar\omega = E_i + E_f$ due to the interband transition being limited by the Fermi sea of excited states. This implies that GP emission from electrical excitation can be more energetic than photon emission from a similar process (that is limited already by $\hbar\omega \lesssim 2E_f$). Finite temperature will broaden all cutoffs by the expected Fermi-Dirac distribution. However, for most frequencies, the GP losses are a more significant source of broadening.

To incorporate the GP losses (as we do in all the figures) we modify the matrix elements calculation by including the imaginary part of the GP wavevector $q_I = q_I(\omega)$, derived independently for each point of the GP dispersion curve [19]. This is equivalent to replacing the delta functions in Eq.2 by Lorentzians with $1/\gamma$ width, defining $\gamma(\omega) = q_R(\omega)/q_I(\omega)$. The calculation can be done partly analytically yielding:

$$\Gamma_{\omega,\theta} = \frac{\alpha c}{\pi^2 \bar{\epsilon}_r v_p(\omega)} \left|\frac{E_i}{\hbar\omega} - 1\right| \int_0^{2\pi} d\varphi \begin{cases} \cos^2(\theta - \varphi/2) & \text{intraband transition} \\ \sin^2(\theta - \varphi/2) & \text{interband transition} \end{cases}$$

$$\cdot \frac{\left|\frac{\sin(\theta)}{\gamma(\omega)}\right|}{\left(\frac{v_p(\omega)}{v_f}\left|\frac{E_i}{\hbar\omega} - 1\right|\sin(\varphi) + \sin(\theta)\right)^2 + \left|\frac{\sin(\theta)}{\gamma(\omega)}\right|^2} \quad (4)$$

$$\cdot \frac{|\cos(\theta)/\gamma(\omega)|}{\left(\frac{v_p(\omega)}{v_f}\left|\frac{E_i}{\hbar\omega} - 1\right|\cos(\varphi) + \cos(\theta) - \frac{v_p(\omega)}{v_f}\frac{E_i}{\hbar\omega}\right)^2 + |\cos(\theta)/\gamma(\omega)|^2}$$

The immediate effect of the GP losses is the broadening of the spectral features, as shown in Figs.2c,3c&4c. Still, the complete analytic theory of Eqs.2a&b matches very well with the exact graphene ČE (e.g., regimes of enhanced emission agree with Eq.3a, as marked in Figs.2b,3b by blue dashed curves). The presence of GP loss also opens up a new regime of quasi-ČE that takes place when the charge velocity is very close to the Čerenkov threshold but does not exceed it. The addition of Lorentzian broadening then closes the gap, creating significant non-zero matrix elements that can lead to *intraband* GP emission (Fig.4). This GP emission occurs even for hot electrons (holes) in positively (negatively) doped graphene, with the only change in Fig.4 being that the upper frequency cutoff is instead shifted to $\hbar\omega \leq E_i - E_f$ (eliminating all interband transitions). The dip in the spectrum at the boundary between interband and intraband transitions (Fig.4c) follows from the charge carriers density of states being zero at the tip of the Dirac cone.

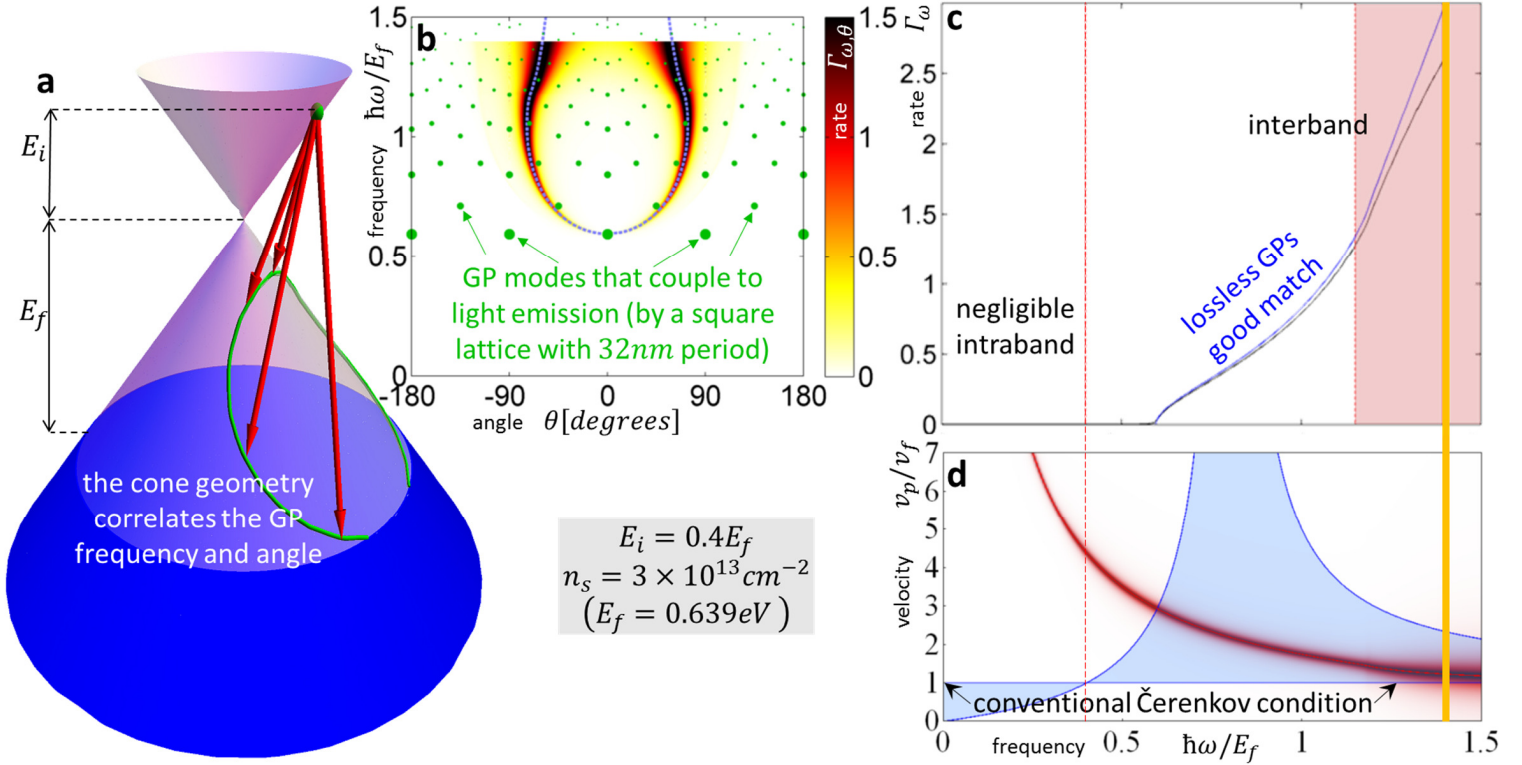

**Figure 3: GP emission from hot carriers.** Caption same as Fig.2. The green dots in (b) show the GPs can be coupled out, as light, with the size illustrating the strength of the coupling.

The interband ČE in Fig.4 shows the possibility of emission of relatively high frequency GPs, even reaching near-infrared and visible frequencies. These are interband transitions as in Figs.2,3 thus limited by $\hbar\omega \leq E_i + E_f$. This limit can get to a few eVs because $E_i$ is controlled externally by the mechanism creating the hot carriers (e.g., p-n junction, tunneling current in a heterostructure, STM tip, ballistic transport in graphene with high drain-source voltage, photoexcitation). Current direct and indirect experimental evidence already shows the existence of GPs at near-infrared frequencies [36-39]. The only fundamental limitation is the energy at which the graphene dispersion ceases to be conical (~$1eV$ from the Dirac point [40]). Even then, our equations are only modified by changing the dispersion relations of the charge carrier and the GP, and therefore the graphene ČE should appear for $E_i$ as high as ~$3eV$ [41]. The equations we presented are still valid since they are written for a general dispersion relation, with $v_p(\omega)$ and $\gamma(\omega)$ as parameters, thus the basic predictions of the equations and the ČE features we describe will continue to hold regardless of the precise plasmon dispersion. For example, a recent paper [42] suggests an alternative way of calculating GP dispersion, giving larger GP phase velocities at high frequencies – this will lead to more efficient GP emission, as well as another intraband regime that can occur without being mediated by the GP loss.

There exist several possible avenues for the observation of the quantum ČE in GPs, having to do with schemes for exciting hot carriers. For example, apart from photoexcitation, hot carriers have been excited from tunneling current in a heterostructure [43], and by a biased STM tip [36], therefore, GPs with the spectral features we predict here (Figs.2c,3c,4c) should be achievable in all these systems. In case the hot carriers are directional, measurement of the GP Čerenkov angle (e.g. Figs.2b,3b,4b) should also be possible. This might be achieved by strong drain-source voltage applied on a graphene p-n junction [44], or in other graphene devices showing ballistic transport [18].

Importantly, the ČE emission of GPs can be coupled out as free-space photons by creating a grating or nanoribbons – fabricated in the graphene, in the substrate, or in a layer above it (e.g., references [37,45-48]) – with two arbitrarily-chosen examples marked by the green dots in Figs.3b,4b. Careful design of the coupling mechanism can restrict the emission to pre-defined frequencies and angles, with further optimization needed for efficient coupling. This clearly indicates that the GP emission, although usually considered as merely a virtual process, can be in fact completely real in some regimes, with the very tangible consequences of light emission in terahertz, infrared or possibly visible frequencies. Such novel sources of light could have promising applications due to graphene's dynamic tunability and small footprint (due to the small scale of GPs). Moreover, near perfect conversion efficiency of electrical energy into photonic energy might be achievable due to the ČE emission rate dominating all other scattering processes. Lastly, unlike plasmonic materials such as silver and gold, graphene is especially exciting in this context since it is CMOS compatible. Still, further research is needed in the design of gratings and/or cavities to minimize losses in the GP-to-photon conversion.

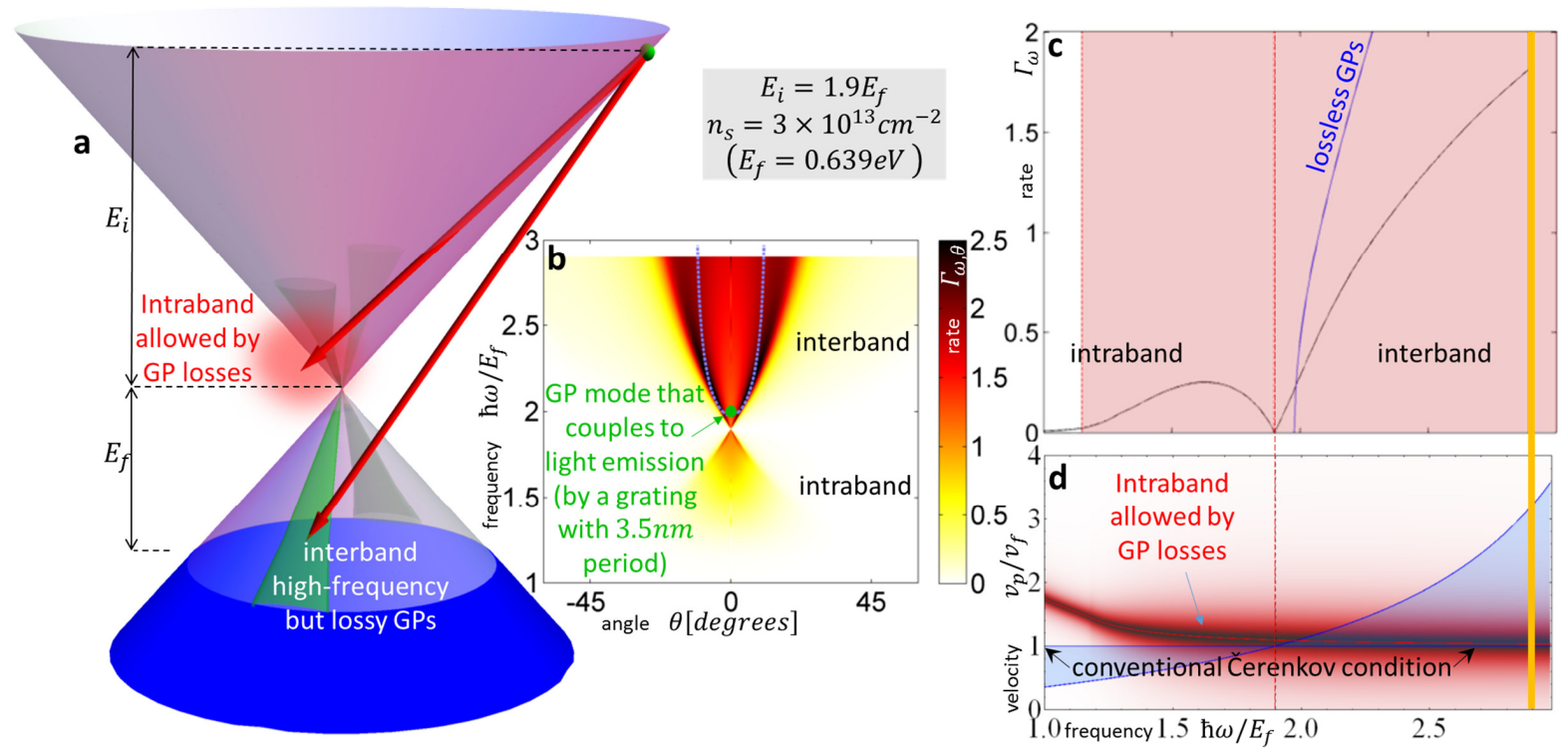

**Figure 4: GP emission from hot carriers.** Caption same as Fig.2. Unlike conventional ČE, most of the emission occurs in the forward direction with a relatively low angular spread. The green dot shows that GPs a particular frequency can be coupled out as light.

The hot carrier lifetime due to GP emission in doped graphene is defined by the inverse of the total rate of GP emission (integrating Eqs.3b/4), and can therefore be exceptionally short (down to a few $fs$). This makes GP emission the dominant decay process (phonon scattering lifetime, in comparison, is on the order of hundreds of $fs$ [7]). Such short lifetimes are in general agreement with previous results (e.g., [49-52]), which state that the ultrashort lifetime of hot carriers in graphene is due to coupling to virtual plasmons (which are part of the electron-electron interaction). The high rates of GP emission also agree with research of the reverse process – of plasmons enhancing and controlling the emission of hot carriers – that is also found to be particularly strong in graphene [53-55]. This might reveal new relations between ČE to other novel ideas of graphene-based radiation sources that are based on different physical principles [56-59].

It is also worth noting that Čerenkov-like plasmon excitations from hot carriers can be found in other condensed matter systems such as a 2D electron gas at the interface of semiconductors. Long before the discovery of graphene, such systems have demonstrated very high Fermi velocities (even higher than graphene's), while also supporting meV plasmons that can have slow phase velocities, partly due to the higher refractive indices possible in such low frequencies [60]. The ČE coupling, therefore, should not be unique to graphene. In many cases [61,62], the coupling of hot carriers to bulk plasmons is even considered as part of the self-energy of the carriers, although the plasmons are then considered as virtual particles in the process. Nonetheless, graphene offers a unique opportunity where the Čerenkov velocity matching can occur at relatively high frequencies, with plasmons that have relatively low losses. Crucially, these differences are what makes the efficiency of the graphene ČE so high. Continued research into other 2D materials (e.g., 2D silver [7]) may lead to materials with higher frequency, lower loss, and higher confinement (lower phase velocity), than graphene plasmons. The prospect of higher frequency plasmons is especially exciting since the ČE radiation intensity increases with frequency (explaining the bluish color of conventional ČE).

We should like to conclude with some very intriguing yet at this stage admittedly very speculative comments. Effects associated with the highly efficient emission of interband GPs and the unexpected emission of intraband GPs predicted by our quantum ČE theory may have already manifested themselves in current graphene experiments, even in ones that do not involve any optical measurement, such as transistor-based graphene devices [20,43]. For example, such GP emission could be a contributing factor to the effect of current saturation observed in graphene devices [20], since large source-drain voltages can take graphene out of equilibrium and create hot carriers. When these hot carriers cross the energy threshold for significant GPs emission they lose energy abruptly, causing a sudden increase in resistivity.     As another example, our graphene ČE might play a role in explaining the surprisingly high frequency of emitted light from graphene [17,21,22], since GPs can couple out as light emission by surface roughness, impurities, etc. This hypothesis is encouraged by reports [17] in which the measurements show characteristics typical of ČE variants, like threshold values and power scaling behavior that do not fit simple black body models. If our theory is indeed applicable here, then the extremely high temperature estimates of the electron gas would need to be modified to account for the contribution of GP emission in the high frequency range of the observed spectrum. This would imply a lower black body radiation spectrum and thus lower graphene temperatures than otherwise expected [22]. Finally, since the GP energy can be higher than both $E_i$ and $E_f$, the ČE could form part of the explanation for the observed frequency up-conversion [21], especially given that multi-plasmon effects are expected due to the high rate of the emission process. Of course, future detailed studies of the systems will be needed to verify the ČE connections proposed here.

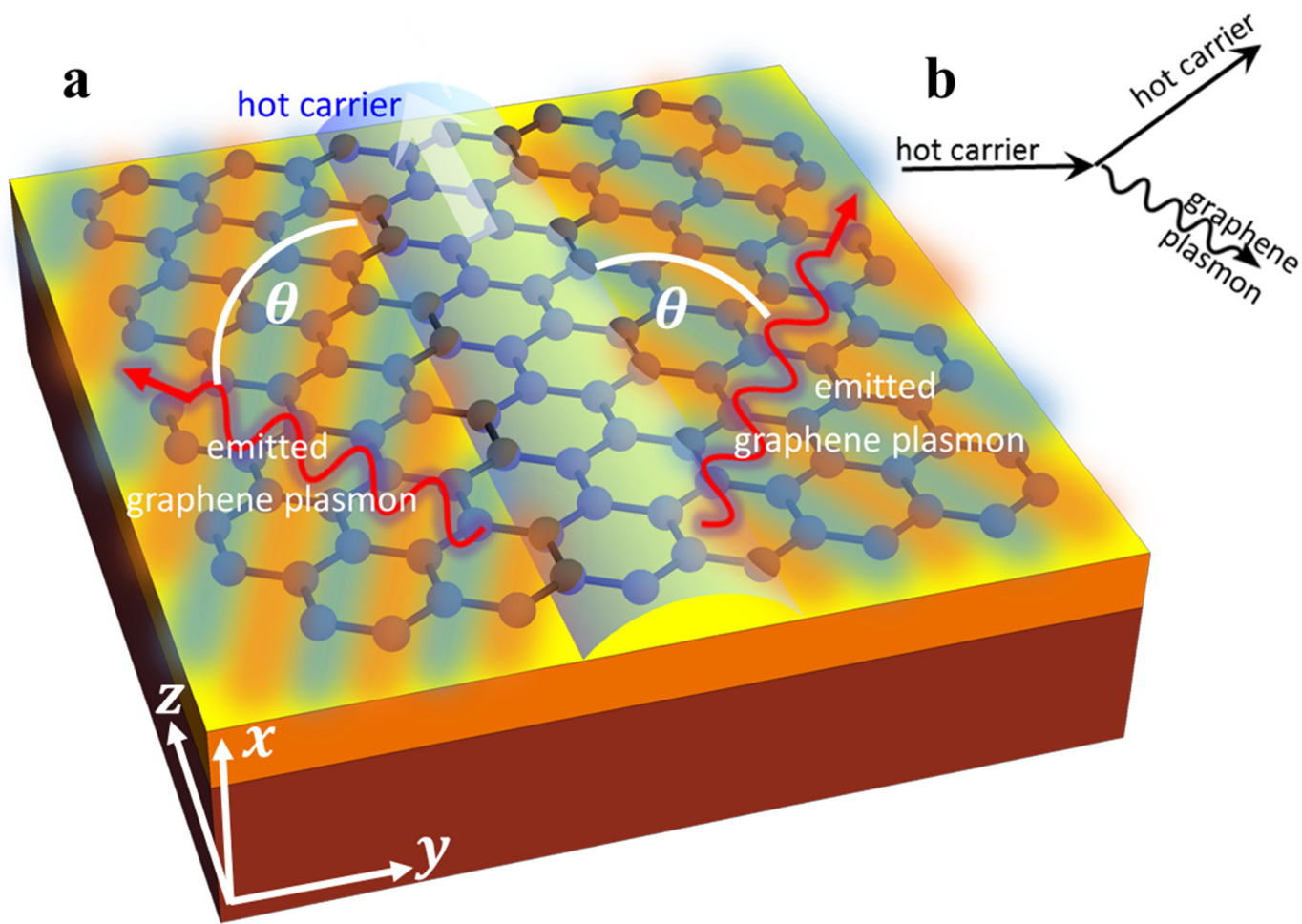

**Figure 1: Illustration of the plasmon emission from charge carriers in graphene via a 2D Čerenkov process.** (a) GP emission in graphene from a hot carrier flowing inside it. The Čerenkov angle into which the GPs are emitted is denoted by $\theta$. (b) A diagram describing the GP emission process from a hot carrier in graphene.

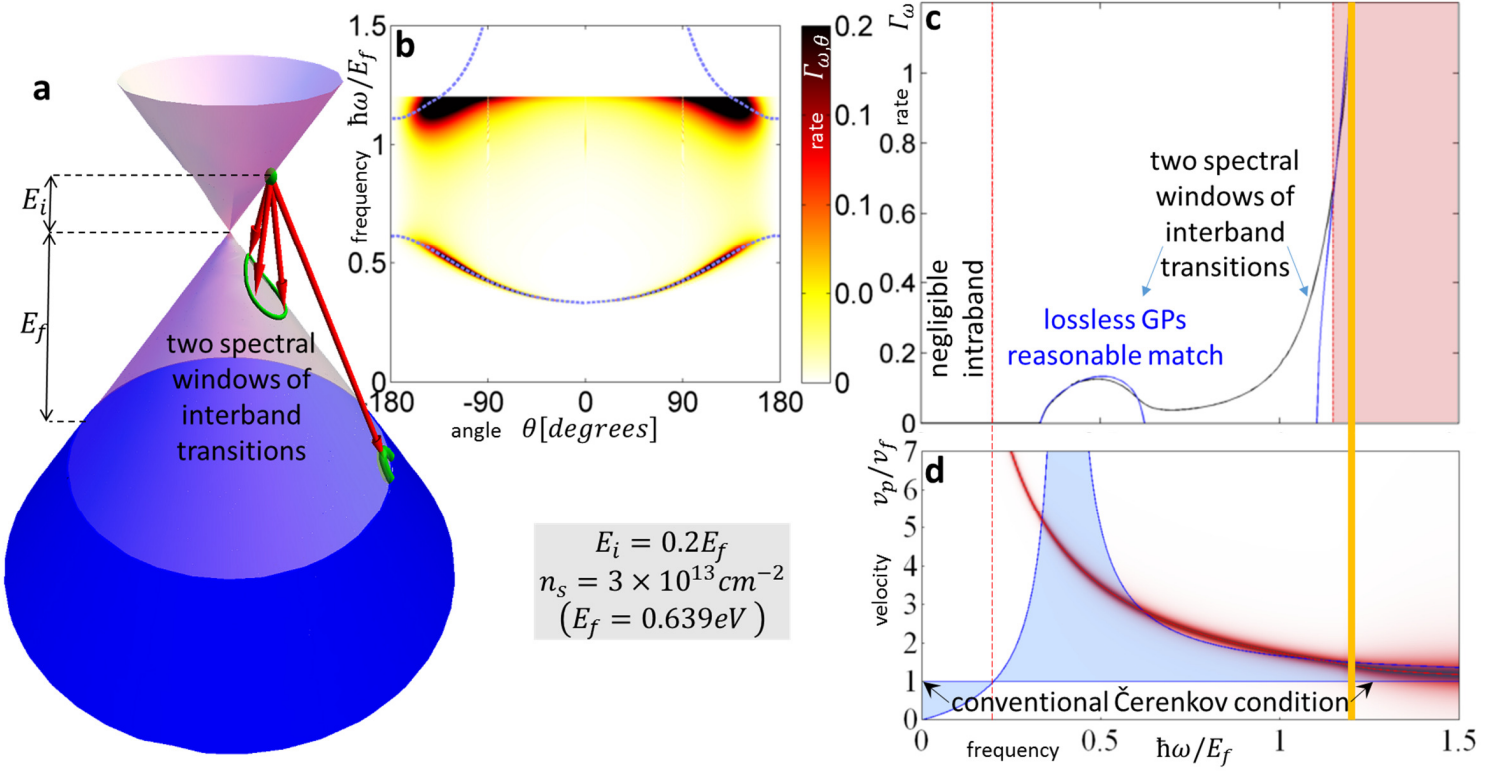

**Figure 2: GP emission from hot carriers.** (a) Illustration of the possible transitions. (b) Map of GP emission rate as a function of frequency and angle, Eq.4. We find most of the GP emission around the dashed blue curves that are exactly found by the Čerenkov angle Eq.3a. (c) Spectrum of the ČE GP emission process, with the red regime marking the area of high losses (as in [19]), the vertical dotted red line dividing between interband to intraband transitions, and the thick orange line marking the spectral cutoff due to the Fermi sea beyond which all states are occupied. Black – emission spectrum with GP losses, Eq.4. Blue – lossless emission approximation, Eq.3. (d) Explaining the GP emission with the quantum ČE. The red curve shows the GP phase velocity, with its thickness illustrating the GP loss. The blue regime shows the range of allowed velocities according to the quantum ČE. We find enhanced GP emission in the frequencies for which the red curve crosses the blue regime, either directly or due to the curve thickness. All figures are presented in normalized units except for the angle shown in degrees.

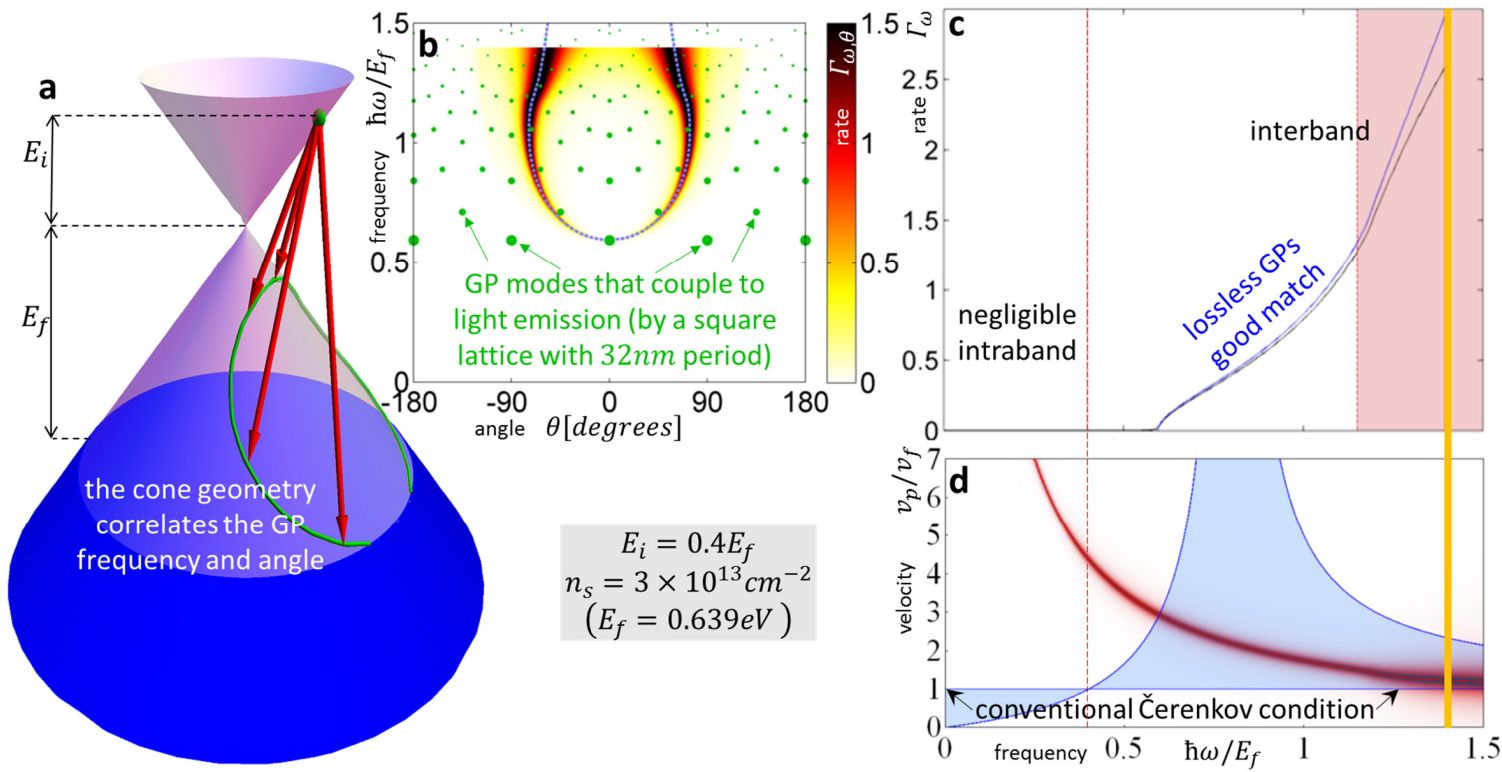

**Figure 3: GP emission from hot carriers.** Caption same as Fig.2. The green dots in (b) show the GPs can be coupled out, as light, with the size illustrating the strength of the coupling.

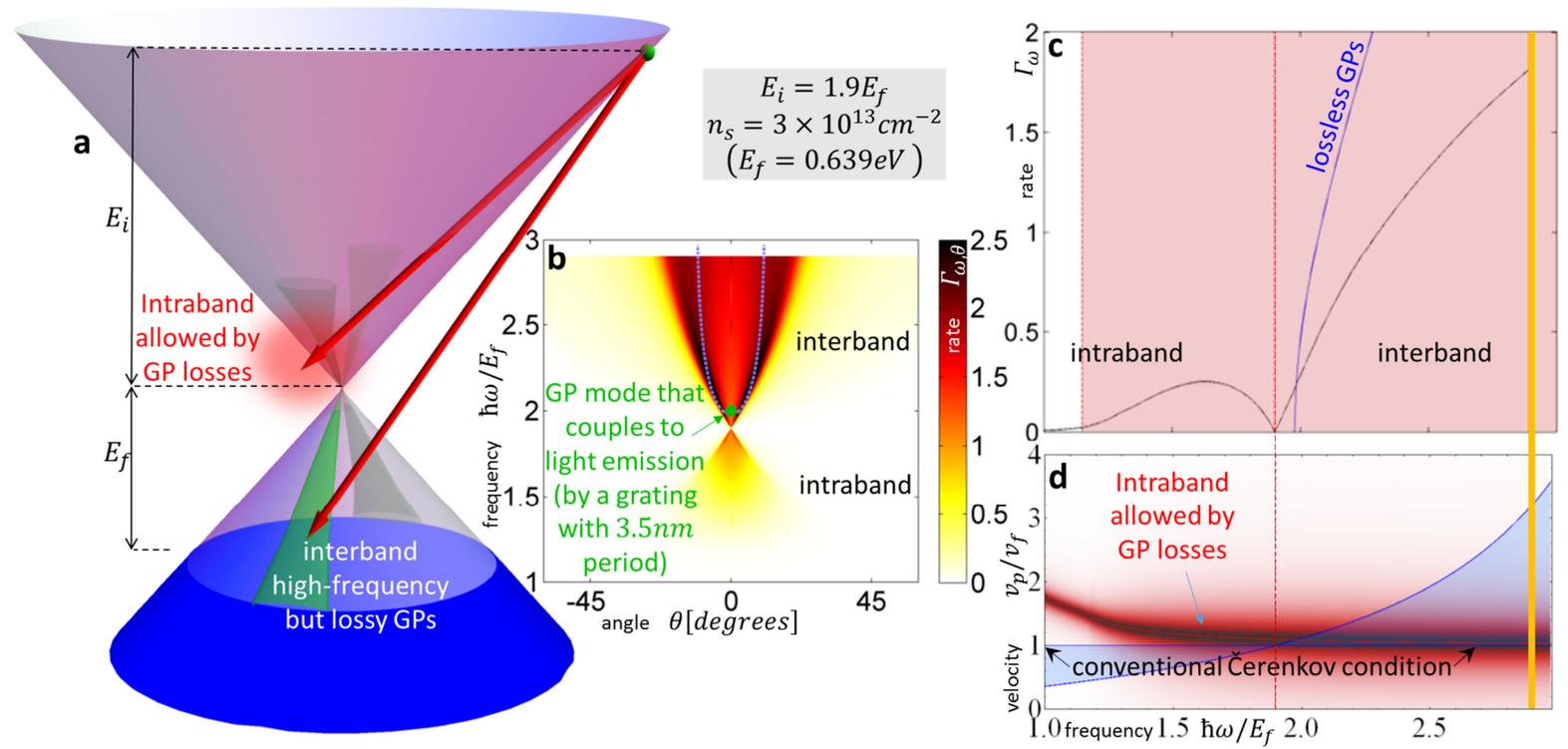

**Figure 4: GP emission from hot carriers.** Caption same as Fig.2. Unlike conventional ČE, most of the emission occurs in the forward direction with a relatively low angular spread. The green dot shows that GPs a particular frequency can be coupled out as light.